\documentclass[11pt,a4paper]{article}
\usepackage{jheppub}

\title{A Reference Model\\ for Anomalously Interacting Bosons}
\author[a,b]{M. V. Chizhov}
\affiliation[a]{CSRT, Faculty of Physics, Sofia University, 1164
Sofia, Bulgaria} \affiliation[b]{DLNP, Joint Institute for Nuclear
Research, 141980, Dubna, Russia}

\emailAdd{mih@phys.uni-sofia.bg}

\abstract{A simple reference model for anomalously interacting
bosons is proposed and implemented in the CompHEP package. This
allows preparing for an experimental search of these bosons at
powerful colliders, such as Tevatron and LHC. New signatures and
some experimental consequences are shortly considered.}

 \arxivnumber{1005.4287}

\toccontinuoustrue
\begin{document}
\maketitle

\flushbottom

\section{Introduction}
The hypothetical heavy neutral bosons, $Z'$, interacting with the
known fermions, quark and leptons, should have a clear signature at
the hadron colliders. Due to their resonance production a peak in
the dilepton invariant high-mass distribution of the Drell--Yan
process can be expected. In the case of leptophobic bosons their
decay in the heavy quarks, $t$ and $b$, allowing tagging and almost
a full kinematics reconstruction, can be investigated.

The extra gauge bosons, $Z'$, occur in any extensions of the
Standard Model (SM) gauge group. If such extensions are motivated by
the Hierarchy Problem, they lead to a new physics around TeV
energies not far above the weak scale. Needless to say,
understanding experimental consequences of the latter is of a
fundamental importance.

However, usually only the gauge bosons, which have minimal couplings
with the SM fermions
\begin{equation}\label{LL}
    {\cal L}_{Z'}=\sum_f \left(g^f_{LL}\;\overline{\psi^f_L}\gamma^\mu\psi^f_L
    +g^f_{RR}\;\overline{\psi^f_R}\gamma^\mu\psi^f_R\right) Z'_\mu
    \, ,
\end{equation}
are considered in literature. In the recent paper~\cite{Dvali} it
was shown, that there are at least three different classes of
theories, all motivated by the hierarchy problem, which predict also
the appearance of new spin-1 weak-doublets, $V_{\mu} \, \equiv
(Z^{*}_{\mu}, W_{\mu}^{*})$, with the internal quantum numbers
identical to the SM Higgs doublet.

In contrast to the well-known $Z'$ and $W'$ bosons they possess only
effective anomalous (magnetic moment type) couplings with ordinary
light fermions~\cite{misha}
\begin{equation}
{\cal L}_{V}= {1\over M} \sum_f\left(g^f_{LR} \, \overline{\Psi^f_L}
\sigma^{\mu\nu} \psi^f_R \, D_{[\mu} V_{\nu]} + g^f_{RL}\,
D^\dag_{[\mu}\,
V^\dag_{\nu]}\,\overline{\psi^f_R}\sigma^{\mu\nu}\Psi^f_L \right),
\label{maincoupling}
\end{equation}
where $M$ is the scale of the new physics, $\psi^f_R$ are the
right-handed singlets and $\Psi^f_L$ are the left-handed doublets.
$D_{\mu}$ are the usual $SU(2)_W\times U(1)_Y$ covariant
derivatives, and the obvious group and family indexes are
suppressed.

Although many Beyond the Standard Models (BSM) have been implemented
in Monte Carlo (MC) Generators like PYTHIA, the new bosons are not
yet even assigned to any number within the MC particle numbering
scheme. The aim of this paper is to suggest a simple reference model
and implement it in one of the most reliable among Matrix Element MC
Generators, the CompHEP package~\cite{CompHEP}. This will allow to
perform simple calculations of the processes involving the new
bosons. In agreement with Les Houches Accord ~\cite{interface} the
corresponding generated event files can be passed to PYTHIA for
showering/hadronization and further also for simulation and
reconstruction within the real experimental software framework.

\section{The reference model and its implementation}

In order to construct the simple reference model for experimental
verifications we need to derive the main features of the new spin-1
bosons from a whole variety of possible models. For example, the
doublet structure always introduces new particles in pairs: neutral
$Z^*$ and charged $W^*$ spin-1 bosons. Moreover, due to non-zero
hypercharge of the doublet the minimal particle set consists of the
four spin-1 bosons: the two charged $W^{*\pm}$ states and the two
neutral $CP$-even Re$Z^*$ and $CP$-odd Im$Z^*$ states.

The crucial common feature of all approaches considered in
\cite{Dvali} is, that the lightness of the Higgs doublets is
guaranteed, because they are related to the spin-1 $V_{\mu}$ bosons
by symmetry. It means that each spin-0 Higgs doublet is associated
with the corresponding spin-1 doublet and vice versa. In the
supersymmetric SM extensions at least two Higgs doublets are needed,
which demands introduction of more than one new spin-1 weak-doublet.
In order to prevent the appearance of flavor-changing neutral
currents one doublet should couple only to up-type quarks while the
other should couple to down-type quarks and charged leptons
only~\cite{GW}.

The different doublets are associated with the different vacuum
expectation values of the Higgs doublets and, in general, acquire
different masses. Let's choose them according to the estimations in
\cite{misha}
\begin{equation}\label{masses}
    M_{Z^*_u}\simeq 700~{\rm GeV},\qquad
    M_{Z^*_d}\simeq 1000~{\rm GeV}.
\end{equation}
In contrast to the case of neutral bosons the charged bosons can
mix, which results in an additional mass splitting. So, the mass of
the lightest boson can be estimated as $M_{W^*}\simeq
500$~GeV~\cite{misha}. It will be shown that all properties of the
new bosons can be described by the following independent input
parameters: coupling constant, masses and mixing of the charged
bosons. All input parameters of the model called \_ESM are
summarized in Table~\ref{tab:vars}.
\begin{table}[t]
\caption{Independent parameters of the \_ESM model (varsX.mdl) }
\label{tab:vars}
\begin{center}
\begin{scriptsize}\hspace{-6.3cm}Variables\\
\begin{tabular}{lll}
  \hspace{0.2cm}Name
  & $|$ \hspace{0.2cm}Value\hspace{0.2cm}
  & $|\!>$  \hspace{1cm}Comment\hspace{1.9cm} $<\!|$\\
  \verb'GW'   &$|$\verb'0.652' &$|$\verb'new coupling constant = EE/SW' \\
  \verb'MZd'  &$|$\verb'1000.' &$|$\verb'mass of down-type Z*' \\
  \verb'MZu'  &$|$\verb'700.' &$|$\verb'mass of up-type Z*' \\
  \verb'MWX'  &$|$\verb'500.' &$|$\verb'mass of W*' \\
  \verb'SX'   &$|$\verb'0.5' &$|$\verb'sin of charged boson mixing'\\
  \hline
  \hline
\end{tabular}
\end{scriptsize}
\end{center}
\end{table}

Table~\ref{tab:prtcls} includes all characteristic representatives
of the \_ESM model and their properties.
\begin{table}[t]
\caption{List of particles of the \_ESM model and their properties
(prtclsX.mdl) } \label{tab:prtcls}
\begin{scriptsize}\hspace{0.3cm}Particles\\
\begin{tabular}{llllllllll}
  Full  name & $|$ P
  & $|$aP\hspace{-0.3cm} & $|$2*spin\hspace{-0.3cm} & $|$mass\hspace{-0.3cm}
  & $|$width\hspace{-0.3cm} & $|$color\hspace{-0.3cm} & $|$aux\hspace{-0.3cm}
  & $|\!>$  \hspace{0.2cm}LaTeX(A)\hspace{0.2cm} $<\!$
  & $|\!>$  \hspace{0.2cm}LaTeX(A+)\hspace{0.2cm} $<\!|$ \\
  \verb'down-type Z*' &$|$\verb'Zd' &$|$\verb'Zd' &$|$\verb'2'
  &$|$\verb'MZd' &$|$\verb'wZd' &$|$\verb'1'
  &$|$ &$|$\verb'Z^*_d' &$|$\verb'Z^*_d' \\
  \verb'up-type ReZ*' &$|$\verb'Zr' &$|$\verb'Zr' &$|$\verb'2'
  &$|$\verb'MZu' &$|$\verb'wZr' &$|$\verb'1'
  &$|$ &$|$\verb'ReZ^*_u' &$|$\verb'ReZ^*_u' \\
  \verb'up-type ImZ*' &$|$\verb'Zi' &$|$\verb'Zi' &$|$\verb'2'
  &$|$\verb'MZu' &$|$\verb'wZi' &$|$\verb'1'
  &$|$ &$|$\verb'ImZ^*_u' &$|$\verb'ImZ^*_u' \\
  \verb'W*' &$|$\verb'X+' &$|$\verb'X-' &$|$\verb'2'
  &$|$\verb'MWX' &$|$\verb'wWX' &$|$\verb'1'
  &$|$ &$|$\verb'W^{*+}' &$|$\verb'W^{*-}' \\
  \hline
  \hline
\end{tabular}
\end{scriptsize}
\end{table}

\noindent Their interactions can be derived from
(\ref{maincoupling})
\begin{eqnarray}\label{L}
    {\cal L}_{ref}&=&\frac{g}{\sqrt{2}M_{Z^*_d}}\left(
    \bar{d}\sigma^{\mu\nu}d+\bar{e}\sigma^{\mu\nu}e
    \right)\partial_\mu {\color{red}Z^*_d}_\nu+
    \frac{\sqrt{2}g}{\sqrt{3}M_{Z^*_u}}\left(\bar{u}\sigma^{\mu\nu}\!u\;
    \partial_\mu {\rm Re}{\color{red}Z^*_u}_\nu+
    i\,\bar{u}\sigma^{\mu\nu}\gamma^5\!u\;
    \partial_\mu {\rm Im}{\color{red}Z^*_u}_\nu\right) \nonumber\\
    &&\hspace{-0.8cm}+\frac{g}{M_{W^*}}\left(\sin\theta_X\,\overline{u_L}\sigma^{\mu\nu}d_R+
    \frac{2}{\sqrt{3}}\cos\theta_X\,\overline{u_R}\sigma^{\mu\nu}d_L+
    \sin\theta_X\,\overline{\nu_L}\sigma^{\mu\nu}e_R
    \right)\partial_\mu {\color{red}W^*_X}^+_\nu + {\rm h.c.},
\end{eqnarray}
where for simplicity the scale of the new physics is equal to the
mass of the corresponding boson and $g$ being the coupling constant
of the $SU(2)_W$ weak gauge group. The flavor universality is
assumed and the family indexes are omitted. The coupling constants
are chosen in such a way that all fermionic decay widths of the new
bosons are the same in the Born approximation and in the massless
quarks limit. This value coincides with the total decay width of the
heavy $W'$ boson of the same mass.
The corresponding decay widths are presented in Table~\ref{tab:func}
taking into account the top quark mass only.
\begin{table}[t]
\caption{Parameters depending on the basic ones (funcX.mdl) }
\label{tab:func}
\begin{scriptsize}\hspace{0.3cm}Constraints\\
\begin{tabular}{lll}
  Name
  & $|\!>$ Expression\hspace{5.2cm} $<\!$
  & $|\!>$ Comment\hspace{2.8cm} $<\!|$\\
  \verb'CX'   &$|$\verb'sqrt(1-SX^2)' &$|$\verb'cos of charged boson mixing' \\
  \verb'pi'   &$|$\verb'acos(-1)' &$|$\verb'Pi' \\
  \verb'Sqrt3' &$|$\verb'sqrt(3)' &$|$\verb'sqrt of 3' \\
  \verb'rtZ'  &$|$\verb'Mtop^2/MZu^2' &$|$\verb'(Mtop/MZu)^2 ratio'
  \\
  \verb'rtW'  &$|$\verb'Mtop^2/MWX^2' &$|$\verb'(Mtop/MWX)^2 ratio'
  \\
  \verb'wZd'  &$|$\verb'GW^2*MZd/(4*pi)' &$|$\verb'width of down-type Z*' \\
  \verb'wZr'
  &$|$\verb'GW^2*MZu/(12*pi)*(2+(1+8*rtZ)*sqrt(1-4*rtZ))'
  &$|$\verb'width of up-type ReZ*'\\
  \verb'wZi'
  &$|$\verb'GW^2*MZu/(12*pi)*(2+sqrt((1-4*rtZ)^3))'
  &$|$\verb'width of up-type ImZ*'\\
  \verb'wWX'
  &$|$\verb'GW^2*MWX/(4*pi)*(1+SX^2*rtW^2*(3-2*rtW)/12)'
  &$|$\verb'width of W*'\\
  \hline
  \hline
\end{tabular}
\end{scriptsize}
\end{table}

The present paper discusses only the resonance production of the
heavy bosons and their subsequent decay into fermion pairs, which is
important for early discoveries. In the case of the light fermions
it is impossible to discriminate the multiplicative quantum numbers
of the neutral boson, $P$ and $C$, because they have identical
signatures. Therefore, we will consider only one of the down-type
neutral bosons, for instance, Re$Z^*_d$. Table~\ref{tab:lgrng}
describes the vertex structures of the new interactions (\ref{L}).
\begin{table}[t]
\caption{Vertices of interactions of the \_ESM model
(lgrngX.mdl) } \label{tab:lgrng}
\begin{scriptsize}\hspace{0.3cm}Lagrangian\\
\begin{tabular}{llllll}
  P1 & $|$P2 & $|$P3 & $|$P4\hspace{-0.3cm}
  & $|\!>$  Factor\hspace{1.3cm} $<\!$
  & $|\!>$ dLagrangian/ dA(p1) dA(p2) dA(p3) \hspace{2.3cm} $<\!|$ \\
  \verb'B' &$|$\verb'b' &$|$\verb'Zd' &$|$
  &$|$\verb'GW/(2*Sqrt2*MZd)' &$|$\verb'G(m3)*G(p3)-G(p3)*G(m3)' \\
  \verb'D' &$|$\verb'd' &$|$\verb'Zd' &$|$
  &$|$\verb'GW/(2*Sqrt2*MZd)' &$|$\verb'G(m3)*G(p3)-G(p3)*G(m3)' \\
  \verb'E' &$|$\verb'e' &$|$\verb'Zd' &$|$
  &$|$\verb'GW/(2*Sqrt2*MZd)' &$|$\verb'G(m3)*G(p3)-G(p3)*G(m3)' \\
  \verb'L' &$|$\verb'l' &$|$\verb'Zd' &$|$
  &$|$\verb'GW/(2*Sqrt2*MZd)' &$|$\verb'G(m3)*G(p3)-G(p3)*G(m3)' \\
  \verb'M' &$|$\verb'm' &$|$\verb'Zd' &$|$
  &$|$\verb'GW/(2*Sqrt2*MZd)' &$|$\verb'G(m3)*G(p3)-G(p3)*G(m3)' \\
  \verb'S' &$|$\verb's' &$|$\verb'Zd' &$|$
  &$|$\verb'GW/(2*Sqrt2*MZd)' &$|$\verb'G(m3)*G(p3)-G(p3)*G(m3)' \\
  \verb'C' &$|$\verb'c' &$|$\verb'Zr' &$|$
  &$|$\verb'GW/(Sqrt2*Sqrt3*MZu)' &$|$\verb'G(m3)*G(p3)-G(p3)*G(m3)' \\
  \verb'T' &$|$\verb't' &$|$\verb'Zr' &$|$
  &$|$\verb'GW/(Sqrt2*Sqrt3*MZu)' &$|$\verb'G(m3)*G(p3)-G(p3)*G(m3)' \\
  \verb'U' &$|$\verb'u' &$|$\verb'Zr' &$|$
  &$|$\verb'GW/(Sqrt2*Sqrt3*MZu)' &$|$\verb'G(m3)*G(p3)-G(p3)*G(m3)' \\
  \verb'C' &$|$\verb'c' &$|$\verb'Zi' &$|$
  &$|$\verb'GW/(Sqrt2*Sqrt3*MZu)' &$|$\verb'(G(m3)*G(p3)-G(p3)*G(m3))*G5' \\
  \verb'T' &$|$\verb't' &$|$\verb'Zi' &$|$
  &$|$\verb'GW/(Sqrt2*Sqrt3*MZu)' &$|$\verb'(G(m3)*G(p3)-G(p3)*G(m3))*G5' \\
  \verb'U' &$|$\verb'u' &$|$\verb'Zi' &$|$
  &$|$\verb'GW/(Sqrt2*Sqrt3*MZu)' &$|$\verb'(G(m3)*G(p3)-G(p3)*G(m3))*G5' \\
  \verb'B' &$|$\verb't' &$|$\verb'X-' &$|$ &$|$\verb'GW/(4*Sqrt3*MWX)'
  &$|$\verb'(G(m3)*G(p3)-G(p3)*G(m3))*(Sqrt3*SX*(1-G5)+2*CX*(1+G5))' \\
  \verb'C' &$|$\verb's' &$|$\verb'X+' &$|$ &$|$\verb'GW/(4*Sqrt3*MWX)'
  &$|$\verb'(G(m3)*G(p3)-G(p3)*G(m3))*(Sqrt3*SX*(1+G5)+2*CX*(1-G5))' \\
  \verb'D' &$|$\verb'u' &$|$\verb'X-' &$|$ &$|$\verb'GW/(4*Sqrt3*MWX)'
  &$|$\verb'(G(m3)*G(p3)-G(p3)*G(m3))*(Sqrt3*SX*(1-G5)+2*CX*(1+G5))' \\
  \verb'E' &$|$\verb'nu' &$|$\verb'X-' &$|$
  &$|$\verb'GW/(4*MWX)'
  &$|$\verb'(G(m3)*G(p3)-G(p3)*G(m3))*SX*(1-G5)' \\
  \verb'L' &$|$\verb'nl' &$|$\verb'X-' &$|$
  &$|$\verb'GW/(4*MWX)'
  &$|$\verb'(G(m3)*G(p3)-G(p3)*G(m3))*SX*(1-G5)' \\
  \verb'M' &$|$\verb'nm' &$|$\verb'X-' &$|$
  &$|$\verb'GW/(4*MWX)'
  &$|$\verb'(G(m3)*G(p3)-G(p3)*G(m3))*SX*(1-G5)' \\
  \verb'Ne' &$|$\verb'e' &$|$\verb'X+' &$|$ &$|$\verb'GW/(4*MWX)'
  &$|$\verb'(G(m3)*G(p3)-G(p3)*G(m3))*SX*(1+G5)' \\
  \verb'Nl' &$|$\verb'l' &$|$\verb'X+' &$|$ &$|$\verb'GW/(4*MWX)'
  &$|$\verb'(G(m3)*G(p3)-G(p3)*G(m3))*SX*(1+G5)' \\
  \verb'Nm' &$|$\verb'm' &$|$\verb'X+' &$|$ &$|$\verb'GW/(4*MWX)'
  &$|$\verb'(G(m3)*G(p3)-G(p3)*G(m3))*SX*(1+G5)' \\
  \verb'S' &$|$\verb'c' &$|$\verb'X-' &$|$ &$|$\verb'GW/(4*Sqrt3*MWX)'
  &$|$\verb'(G(m3)*G(p3)-G(p3)*G(m3))*(Sqrt3*SX*(1-G5)+2*CX*(1+G5))' \\
  \verb'T' &$|$\verb'b' &$|$\verb'X+' &$|$ &$|$\verb'GW/(4*Sqrt3*MWX)'
  &$|$\verb'(G(m3)*G(p3)-G(p3)*G(m3))*(Sqrt3*SX*(1+G5)+2*CX*(1-G5))' \\
  \verb'U' &$|$\verb'd' &$|$\verb'X+' &$|$ &$|$\verb'GW/(4*Sqrt3*MWX)'
  &$|$\verb'(G(m3)*G(p3)-G(p3)*G(m3))*(Sqrt3*SX*(1+G5)+2*CX*(1-G5))' \\
  \hline
  \hline
\end{tabular}
\end{scriptsize}
\end{table}

These four tables completely describe the model and can be used for
process evaluations within the CompHEP package.

\section{Consequences for colliders}
The hadron colliders, due to their biggest center-of-momentum (CM)
energy $\sqrt{s}\sim$ several TeVs and their relatively compact
sizes, still remain the main tools for discoveries of very heavy
particles. Since the production mechanism for new heavy bosons at a
hadron collider is the $q\bar{q}$ resonance fusion, the  presence of
partons with a broad range of different momenta allows to flush the
entire energetically accessible region, roughly, up to $\sqrt{s}/2$.
This, in a way, fixes the dominant production and decay mechanisms.

In paper \cite{two} it has been found that tensor interactions lead
to a new angular distribution of the outgoing fermions
\begin{equation}\label{GLR}
    \frac{{\rm d} \sigma(q\bar{q}\to Z^*\!/W^*\to f\bar{f})}
    {{\rm d} \cos\theta} \propto
    \cos^2\theta,
\end{equation}
in comparison with the well-known vector interactions
\begin{equation}\label{GLL}
    \frac{{\rm d} \sigma(q\bar{q}\to Z'\!/W'\to f\bar{f})}
    {{\rm d} \cos\theta} \propto
    1+\cos^2\theta \, .
\end{equation}
It was realized later~\cite{misha} that this property ensures
distinctive signature for the detection of the new interactions at
the hadron colliders. At a first glance, the small difference
between the distributions (\ref{GLR}) and (\ref{GLL}) seems
unimportant. However, the absence of the constant term in the first
case results in new experimental signatures.

\begin{figure}[h]
\epsfig{file=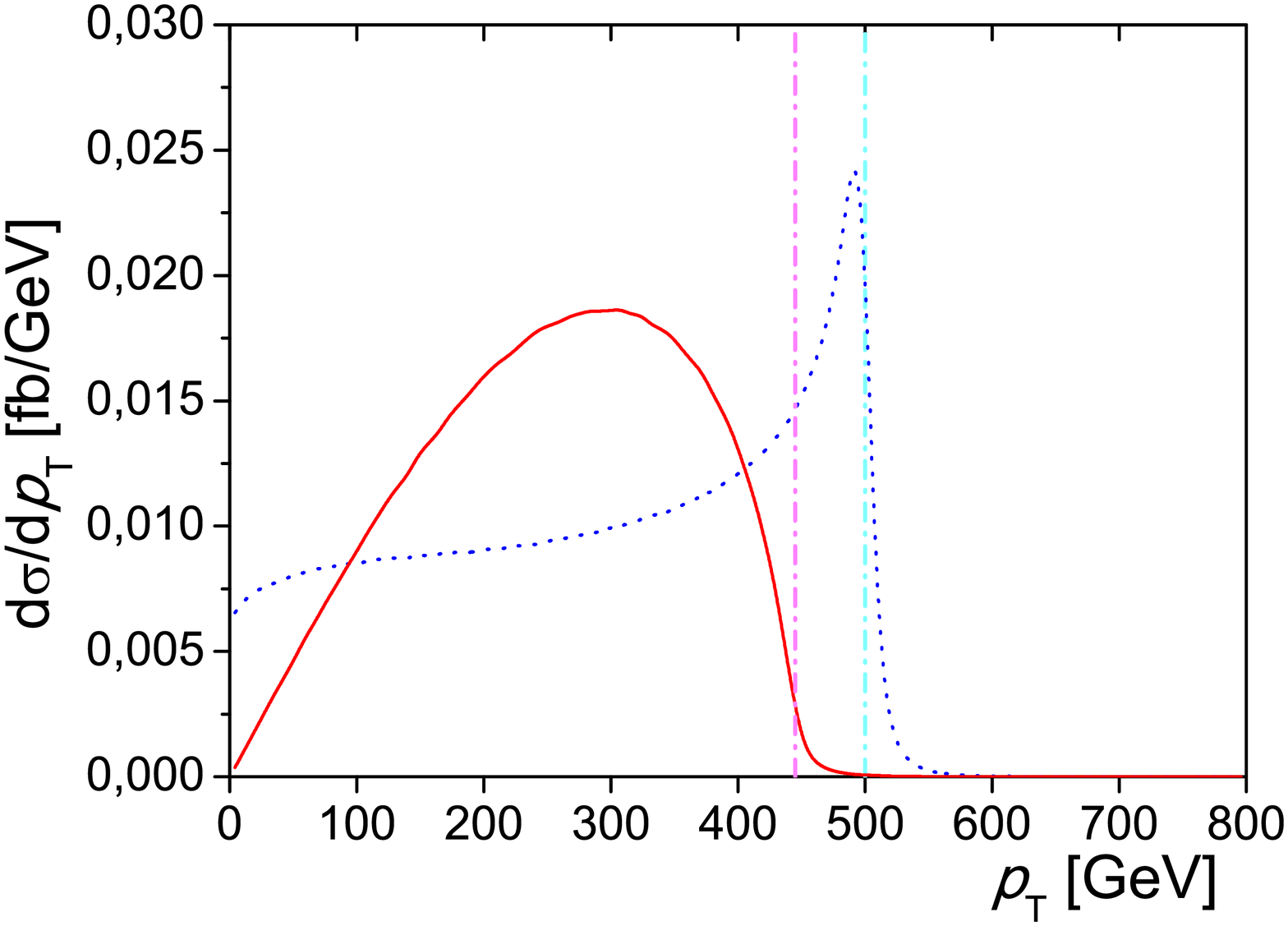,width=0.49\textwidth}
\epsfig{file=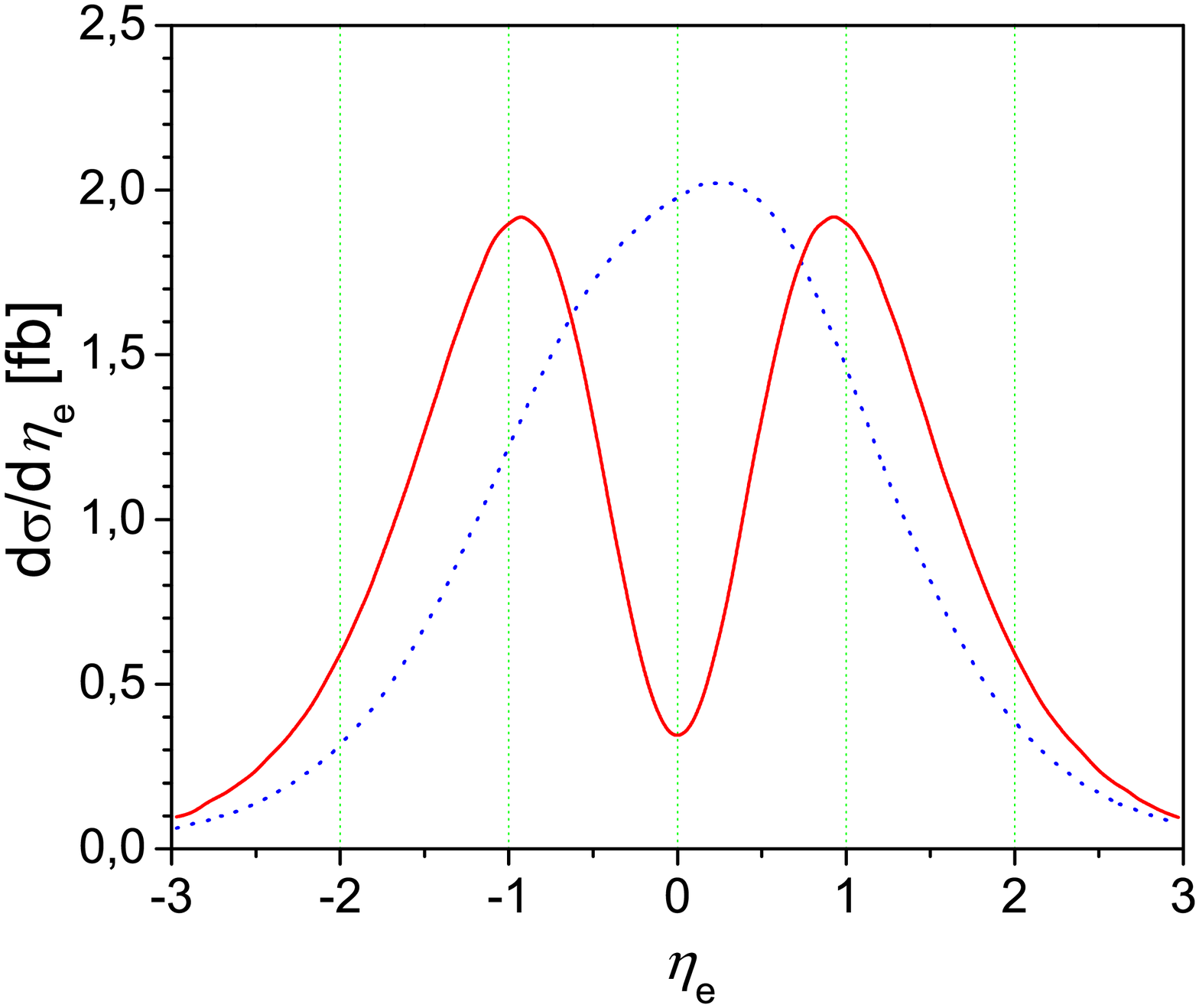,width=0.45\textwidth} \caption{\label{fig:comp}
Left: the final electron $p_{\rm T}$ distributions from the $Z^*_d$
(solid) and $Z'_{SSM}$ (dotted) bosons decays. Right: the
differential cross-sections for the $Z^*$ (solid) and $Z'_{SSM}$
(dotted) bosons as functions of the pseudorapidity of the final
electron. }
\end{figure}

First of all, since the angular distribution for vector interactions
(\ref{GLL}) includes a nonzero constant term, this leads to the
kinematical singularity in $p_{\rm T}$ distribution of the final
fermion
\begin{equation}\label{1/cos}
    \frac{1}{\cos\theta}\propto\frac{1}{\sqrt{(M/2)^2-p^2_{\rm T}}}
\end{equation}
in the narrow width approximation $\Gamma <\!\!\!< M$
\begin{equation}\label{narrow}
    \frac{1}{(s-M^2)^2+M^2\Gamma^2}\approx\frac{\pi}{M\Gamma}\delta(s-M^2).
\end{equation}
This singularity is transformed into the well-known Jacobian peak
due to a finite width of the resonance. In contrast to this, the
pole in the decay distribution of the $Z^*/W^*$ bosons is canceled
out and the fermion $p_{\rm T}$ distribution reaches even zero at
the kinematical endpoint $p_{\rm T}=M/2$. Therefore, the $Z^*/W^*$
boson decay distribution has a broad smooth hump with a maximum
below the kinematical endpoint, instead of a sharp Jacobian peak
(the left-hand plot of Fig.~\ref{fig:comp}).\footnote{All
calculations are performed for the resonant production of the
down-type chiral $Z^*_d$ bosons with the mass $M=890$~GeV and for
the sequential $Z'_{SSM}$ bosons with the mass $M=1000$~GeV and
their subsequent decays into $e^+e^-$ pairs in $p\bar{p}$ collisions
at $\sqrt{s}=1.96$~TeV using the CompHEP package 4.5.1. Thus the
chosen masses lead to the identical cross sections of 5.6~fb.}

According to Eq. (\ref{GLR}), for the $Z^*/W^*$ bosons there exists
a characteristic plane, perpendicular to the beam axis in the parton
rest frame, where the emission of the final-state pairs is
forbidden. The nonzero probability in the perpendicular direction in
the laboratory frame is only due to the longitudinal boosts of the
colliding partons. This property is responsible for the additional
dips in the middle of the final fermion pseudorapidity distributions
of the anomalously interacting bosons, in contrast to the $Z'/W'$
bosons with the minimal gauge couplings (the right-hand plot of
Fig.~\ref{fig:comp}).

\begin{figure}[h]
\epsfig{file=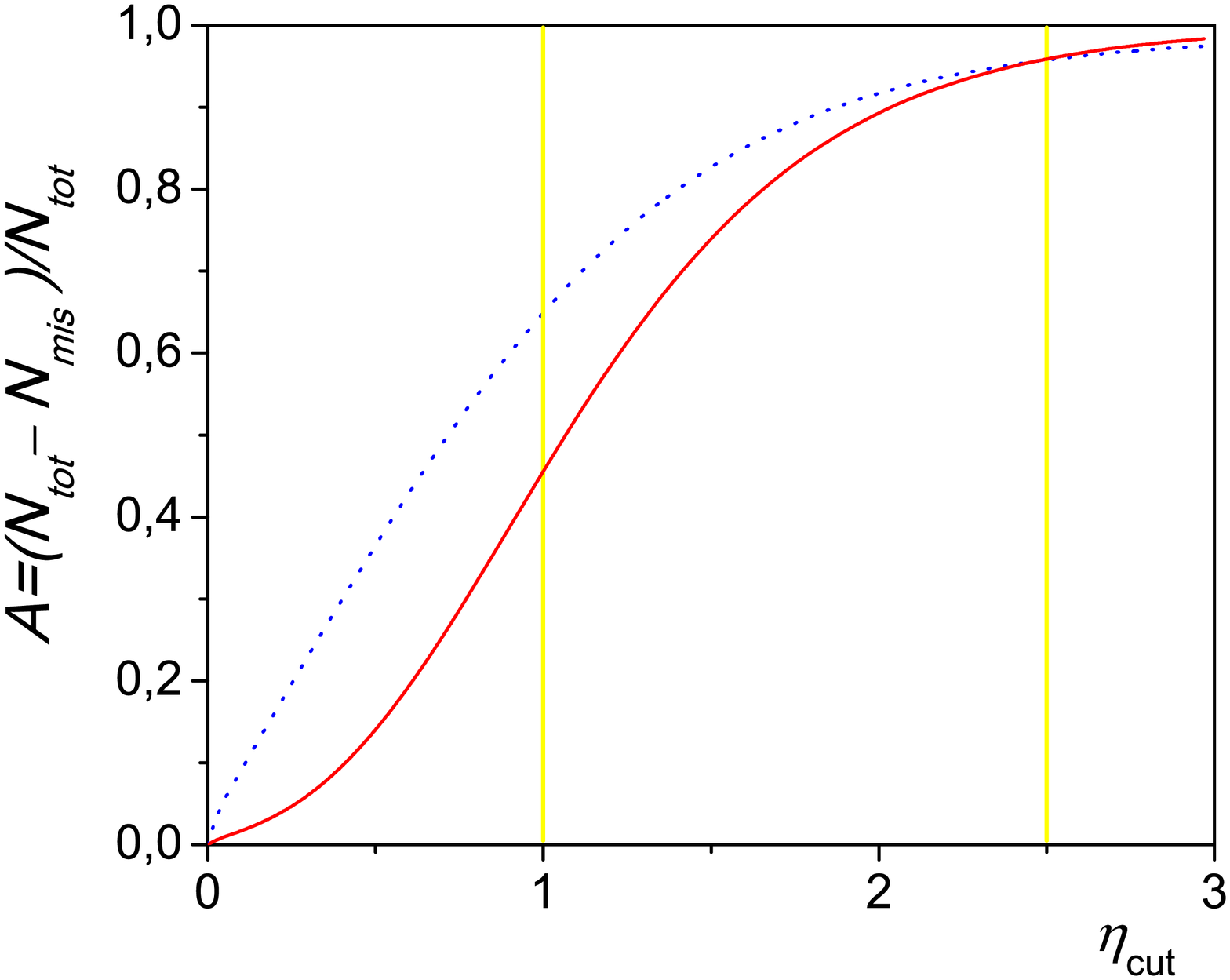,width=0.45\textwidth}
\epsfig{file=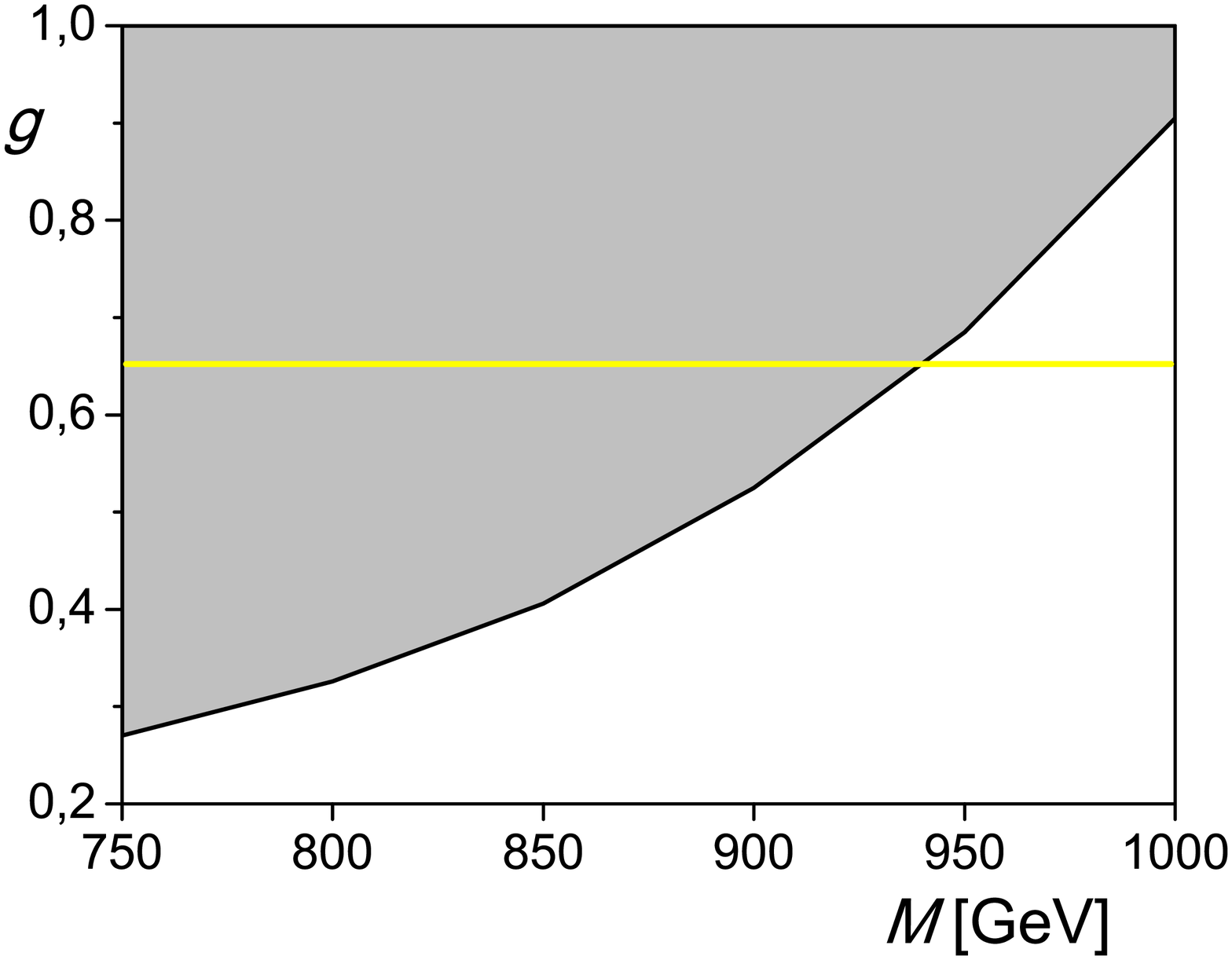,width=0.49\textwidth}
\caption{\label{fig:gM} Left: the acceptances for the $Z^*_d$
(solid) and $Z'_{SSM}$ (dotted) bosons as function of
$|\eta_e|<\eta_{\rm cut}$ cut. Right: the allowed region (white) for
the coupling constant$-$mass relation of the $Z^*_d$ resonance from
the D0 Tevatron constraints~\cite{D0}. The horizontal line
corresponds to the ordinary value of the $SU(2)_W$ coupling constant
$g$.}
\end{figure}

While the maximum of the gauge boson distribution is centered at the
small lepton pseudorapidities, which correspond to the central part
of the detector, the chiral boson distribution has minimum in this
region and its maxima are placed at the edges of the CDF and D0
central calorimeters $|\eta|\simeq 1$. Based on the fact, that the
major part of the leptons, stemming from the $Z'$ decays, are
emitted in the central detector region, both collaborations have
analyzed the spectrum of the transverse high-energy electrons only
in the central electromagnetic calorimeters $|\eta_e|\leq
\eta_{cut}\simeq 1$.
The acceptances for the processes with the chiral $Z^*_d$ bosons and
the sequential $Z'_{SSM}$ bosons are shown in the left-hand plot of
Fig.~\ref{fig:gM}.

The cuts in the backward-forward regions lead to a miss in an
essential part of the events from the chiral boson decays due to the
mentioned previously specific angular distribution. As seen from the
left-hand plot of Fig.~\ref{fig:gM} the curve for the chiral $Z^*_d$
boson mostly lies under the $Z'_{SSM}$ curve, and at $\eta_{cut}
\simeq 1$ there are around 65\% detected events in the case of the
sequential $Z'_{SSM}$ bosons decays and only 45\% in the case of the
chiral $Z^*_d$ bosons decays.

Taking into account this fact we can correct and apply experimental
constraints of the D0 Collaboration (TABLE~II from \cite{D0}) on
production cross section in the case of the chiral $Z^*_d$ bosons
with arbitrary coupling constant $g$ (the right-hand plot of
Fig.~\ref{fig:gM}). One can see that due to a smaller acceptance for
the $Z^*_d$ bosons than for the $Z'_{SSM}$ bosons, even for the same
production cross sections more lighter $Z^*_d$ masses are allowed.
For example, in the case when the coupling constant $g$ is equal to
the ordinary value of the $SU(2)_W$ coupling constant, we obtain
$M_{Z^*_d}>940$~GeV, in comparison with D0 result
$M_{Z'_{SSM}}>1023$~GeV.

\acknowledgments{I am grateful to Patrice Verdier for the very
useful discussions.}


\end{document}